\documentstyle[aabib,epsfig]{l-aa-ps}
\begin{document}
    \newcommand{\rf}{\sf}
    \newcommand{\E}{{\em Einstein}}
    \newcommand{\EMSS}{Extended Medium Sensitivity Survey}
    \newcommand{\ESS} {{\em Einstein} Slew Survey}
    \newcommand{\EX}{{\em EXOSAT }}
    \newcommand{\Mv}{\mbox{$ M_{\rm V} $}}
    \newcommand{\mv}{\mbox{$ m_{\rm V} $}}
    \newcommand{\bv}{\mbox{$ B-V $}}
    \newcommand{\ri}{\mbox{$ R-I $}}
    \newcommand{\by}{\mbox{$ b-y $}}
    \newcommand{\MH}{\mbox{$[{\rm M}/{\rm H}]$}}
    \newcommand{\fxfv}{\mbox{$f_{\rm X}/f_{\rm v}$}}
    \newcommand{\fv}{\mbox{$f_{\rm v}$}}
    \newcommand{\lx}{\mbox{$L_{\rm X}$}}
    \newcommand{\wli}{\mbox{$ W({\rm Li}) $}}
    \newcommand{\nli}{\mbox{$ N({\rm Li}) $}}
    \newcommand{\nh}{\mbox{$ N({\rm H}) $}}
    \newcommand{\cq}{\mbox{$ \chi^2$}}
    \newcommand{\XLF}{X-ray luminosity function}
    \newcommand{\XLFs}{X-ray luminosity functions}
    \newcommand{\vsini}{\mbox{$v \sin(i)$}}
    \newcommand{\teff}{\mbox{$T_{\rm eff}$}}
    \newcommand{\lii}{\mbox{Li\,{\sc i}}}
    \newcommand{\fei}{\mbox{Fe\,{\sc i}}}
    \newcommand{\feii}{\mbox{Fe\,{\sc ii}}}
    \newcommand{\ali}{\mbox{Al\,{\sc i}}}
    \newcommand{\cai}{\mbox{Ca\,{\sc i}}}
    \newcommand{\cri}{\mbox{Cr\,{\sc i}}}
    \newcommand{\feh}{\mbox{[Fe/H]}}
    \newcommand{\lgf}{\mbox{$\log gf$}}
    \newcommand{\logg}{\mbox{$\log(g)$}}
    \newcommand{\dof}{\mbox{degrees of freedom}}
    \newcommand{\muno}{\mbox{$m_1$}}
    \newcommand{\dmuno}{\mbox{$\delta m_1$}}
    \newcommand{\bcet}{\mbox{$\beta$~Cet}}

\thesaurus{6(08.01.1; 08.12.1; 13.25.5; 03.13.2)}

\title{On coronal abundances derived with the SAX/LECS and ASCA/SIS
  detectors}%\thanks{}}

\author{F. Favata\inst{1} 
  \and A. Maggio\inst{2}
  \and G. Peres\inst{2} 
  \and S. Sciortino\inst{2} }

\institute{Astrophysics Division -- Space Science Department of ESA, ESTEC,
 Postbus 299, NL-2200 AG Noordwijk, The Netherlands
\and
 Istituto e Osservatorio Astronomico di Palermo, 
 Piazza del Parlamento 1, I-90134 Palermo, Italy \\
}

\offprints{F. Favata (ffavata@astro.estec.esa.nl)}

\date{Received date 17 April 1997; accepted date}

\maketitle 

%\markboth{F. Favata et al., On the comparison between spectroscopic
%and photometric metallicity}{On the comparison between spectroscopic
%and photometric metallicity}

\begin{abstract}
  
  We have studied the performance of global \cq\ fitting of
  low-resolution X-ray spectra in retrieving intrinsic source
  parameters, with emphasis on the coronal metallicity. The study has
  been conducted by fitting large numbers of simulated spectra with
  known characteristics, and studying the distribution of best-fit
  parameters. We have studied the behavior of the LECS detector
  on board the SAX satellite and the SIS detector on board the ASCA
  satellite. The fitted source spectra have either two discrete
  temperature components or a power-law temperature distribution, with
  metallicity variations modeled by a single global abundance
  parameter. The model used for the fitting has always been a
  two-temperature one, with global varying abundance, to explore the
  influence of the a priori ignorance of the actual temperature
  stratification in the source being observed.

  The simulations performed explore the influence of varying
  statistics in the observed spectrum (spanning a realistic range of
  values) as well as the effect of varying the intrinsic source
  metallicity, with values in the range 0.15--1.0 times the solar
  value. We find that the source metallicity can be retrieved within
  few tens of percent from ASCA/SIS spectra of typical signal to noise
  ratio, and within few percent from SAX/LECS spectra at the same
  signal to noise ratio.  However relatively small uncertainties in
  the detector calibrations and in the plasma emission codes are
  likely to potentially cause large systematic off-sets in the value
  of the best-fit parameters.  Similar systematic off-sets may derive
  from assuming too simplistic a temperature distribution for the
  source plasma.

  In addition we have re-analyzed the ASCA/SIS spectra of the active
  giants \bcet\ and Capella with the same set of assumptions used in
  the simulations, showing how the best-fit metallicity in these two
  real cases depends on the details of the fitting process, and in
  particular on the chosen energy range.


\end{abstract}

\section{Introduction}
\label{sec:intro}

Currently available soft X-ray spectrographs are of the non-dispersive
type (CCDs and GSPCs), and the ``analysis'' of a spectrum, as commonly
performed, consists essentially in searching, given a detector
response matrix and a plasma emission code, the space of possible
``models'' (i.e.\ distributions of temperatures and relative
normalizations, as well as elemental abundances), for the ``best-fit''
one. This is usually done within a set of a priori constraints on the
model, in particular fixing the number of discrete temperature
components allowed. The usual measure of ``goodness of fit'' used is
the \cq, and the fitting process thus consists in a search, through
\cq\ space, for the model yielding the minimum \cq\ value.

One notable result of X-ray coronal astronomy coming from ASCA is that
most (if not all) coronal spectra observed so far with the SIS CCD
detector yield, when fit with currently available plasma emission
codes, best-fit parameters incompatible with solar (or near-solar)
abundance plasmas. At the same time they appear to be ``better fit''
with elemental abundances typically a fraction of the solar abundance,
with values as low as 0.1 times solar not uncommon. This finding
points toward a widespread lower-than-solar metal abundance in stellar
coronae, although departures from the stellar photospheric abundances,
which would be of greater importance, have not been usually
investigated.

The limited spectral resolution of non-dispersive X-ray detectors
makes it impossible to determine reliably the flux of individual lines
or line complexes, so that most studies of coronal abundances up to
now have been conducted by doing global fits to the whole spectrum,
using the resulting best-fit parameters (including the abundances) as
the most probable source parameters. In the presence of complex
spectral models, however, it is difficult to assess from first
principles what the actual uncertainties related to the fitting
process are, and how are the various fit parameters (possibly)
correlated with each other.

To help assess the significance of the abundance estimates derived
from low-resolution X-ray spectra, as well as to help compare the
relative performance of the SIS and LECS instruments for this type of
studies, we have performed an extensive set of simulations, to study
the sensitivity of the two instruments to variations in the
metallicity in the source spectrum, and to assess the relative
uncertainties in the derived source parameters. We have first assumed
that the source spectrum is intrinsically emitted by two distinct
isothermal components, i.e. a so-called ``two-temperature'' model.
This model is bound to be a simplistic representation of a more
complicated reality, with the coronal plasma of real stars certainly
having a much more complex distribution of temperatures, but the
two-temperature model is nevertheless still the work-horse of present
day non-dispersive X-ray coronal spectroscopy. Previous work has shown
that the spectrum of multi-temperature plasmas confined in coronal
loops and yielding a moderate number of counts (as the ones discussed
here) may be reasonably well fit by two-temperature models
(\cite{cmp97}). This evidence justifies, for the purpose of the
present paper, the usage of two-temperature models as
``representative'' of a more complex reality. The two-temperature
source spectra have been fitted with two-temperature models.

As a check for the dependence of the fitting process on the
correctness of the assumed source spectrum we have performed an
additional set of simulations assuming that the source has a power-law
temperature distribution, with spectral index $\alpha = 1.5$ and
maximum temperature of 3.0\,keV. This type of power-law distribution
is reasonably representative, for example, of the emission measure
distribution of the solar corona in the temperature range accessible
to soft X-ray detectors (\cite{rd81}). These power-law spectra were
fit with the same type of two-temperature models as the
two-temperature source spectra, thus simulating our ignorance about
the ``true'' intrinsic source temperature distribution. In the present
work the plasma emission model used has been the {\sc mekal} code
(\cite{mkl95}), as implemented in the XSPEC X-ray spectral analysis
package (version 9.0)

\section{Characteristics of the instruments}

The LECS instrument (\cite{pmb+97}) is a drift-less gas scintillation
proportional counter with a thin entrance window, providing continuous
energy coverage in the spectral range from 0.1 to 10\,keV. Its
resolution is energy-dependent (varying as $E^{-0.5}$), and becomes
comparable to the resolution of the CCD-based SIS detector at $\simeq
0.5$\,keV.

\begin{figure}[thbp]
  \begin{center}
    \leavevmode
%    \picplace{7.0 cm}
    \epsfig{file=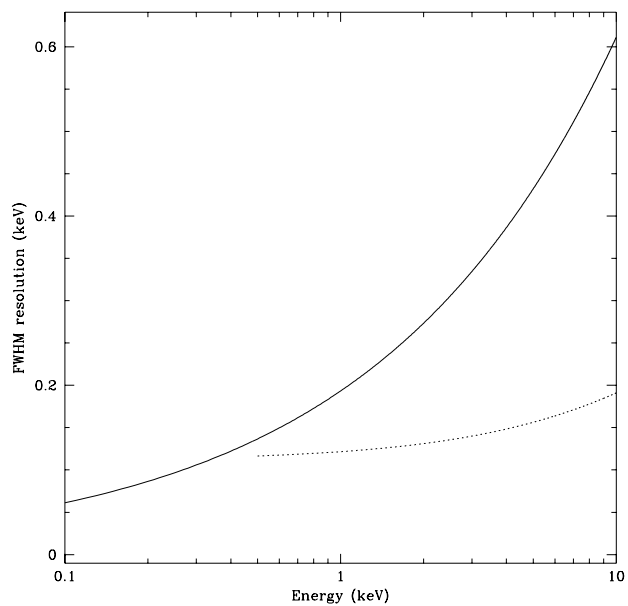, width=8.8cm, bbllx=0pt, bblly=170pt,
      bburx=580pt, bbury=700pt, clip=}
  \end{center}
  \caption{The full width at half maximum (FWHM) spectral resolution
    of the SAX/LECS (continuous line) and ASCA/SIS (dashed line)
    instruments. The SIS resolution is the one at launch time.}
  \label{fig:resolution}
\end{figure}

The SIS instrument is a CCD-based non-dispersive X-ray spectrometer,
providing energy coverage in the band from 0.5 to 10.0\,keV, with an
approximately constant energy resolution, which is effectively
decreasing with time because of the radiation damage of the CCD chip
(\cite{dmy+96}). At the time of the launch the effective resolution
was $\simeq 100$\,eV at 1\,keV, with the precise value depending on
the details of the instrumental mode being used. However, the number
of spectral resolution elements for the SIS and for the LECS is
similar, due to the larger passband of the latter. The spectral
resolution of the LECS and of the SIS (at launch) are compared in
Fig.~\ref{fig:resolution}.  We have used a SIS response appropriate
for an observation made in early 1995, i.e. two years into operations.
This implies a detector performance intermediate between the best
possible performance (at launch) and present-day performance.

\begin{figure}[thbp]
  \begin{center}
    \leavevmode
%    \picplace{7.0 cm}
    \epsfig{file=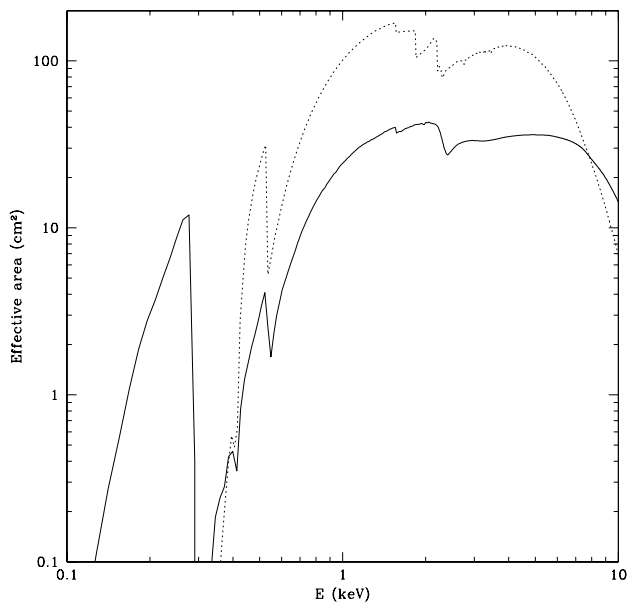, width=8.8cm, bbllx=0pt, bblly=170pt,
      bburx=580pt, bbury=700pt, clip=}
  \end{center}
  \caption{The effective areas of the SAX/LECS (continuous line) and
    ASCA/SIS (dashed line) instruments.}
  \label{fig:area}
\end{figure}

The effective area of the SIS instrument is, thanks to its bigger
mirror, larger than the effective area of the LECS, thus giving it
access to fainter sources. The wider passband of the LECS
partially compensates, for soft source spectra as the one discussed
here, the smaller effective area of the mirrors, so that a given
source with the spectral parameters used here yields a comparable
count rate both in the SIS and in the LECS. The effective
areas of the SIS and of the LECS detector are compared in
Fig.~\ref{fig:area}.

\section{The simulations}

In the first set of simulations a two-temperature source spectrum was
used, with temperatures of 0.5 and 2.0\,keV, and identical emission
measure. Real coronal sources, when modeled with a two-temperature
spectrum, display a wide range of parameters, but the values assumed
here are reasonably representative of ``typical'' active stars, as
determined, for example, by fitting SIS data.  Since the ``true''
coronal source spectra are likely to be produced by a plasma with a
more complex temperature and density distribution, as inferred from
spatially-resolved observations of the solar corona, a second set of
simulations was performed, using a power-law emission measure
distribution with slope $\alpha = 1.5$ and maximum temperature $T_{\rm
  max} = 3$ keV for the source spectrum. These parameters are
comparable to those found by \cite*{pre97} when fitting ROSAT/PSPC
spectra of field solar-type stars.

\begin{table}[thbp]
  \caption{The numerical experiments performed for the present
    paper. Two-temperatures source spectra were generated with
    temperatures of 0.5 and 2.0\,keV, and with an emission measure
    ratio of 1. Power-law spectra have a slope of $-1.5$ and a maximum
    temperature of 3.0\,keV.}
\label{tab:des}
\begin{flushleft}
%\scriptsize
  \footnotesize
\begin{tabular}{rrrc} \hline \\[-2pt]
  Code & counts & Z & Detector \\ \hline\hline
\multicolumn{4}{c}{\em 2-T source spectra} \\[1pt]

  A2TL0.15 & 2500 & 0.15 & ASCA/SIS \\
  A2TM0.15 & 10\,000 & 0.15 & ASCA/SIS \\
  A2TH0.15 & 40\,000 & 0.15 & ASCA/SIS \\
  A2TL0.50 & 2500 & 0.50 & ASCA/SIS \\
  A2TM0.50 & 10\,000 & 0.50 & ASCA/SIS \\
  A2TH0.50 & 40\,000 & 0.50 & ASCA/SIS \\
  A2TL1.00 & 2500 & 1.00 & ASCA/SIS \\
  A2TM1.00 & 10\,000 & 1.00 & ASCA/SIS \\
  A2TH1.00 & 40\,000 & 1.00 & ASCA/SIS \\[1pt]
  S2TL0.15  & 2500 & 0.15 & SAX/LECS \\
  S2TM0.15  & 10\,000 & 0.15 & SAX/LECS \\
  S2TH0.15  & 40\,000 & 0.15 & SAX/LECS \\
  S2TL0.50  & 2500 & 0.50 &  SAX/LECS \\
  S2TM0.50  & 10\,000 & 0.50 & SAX/LECS \\
  S2TH0.50  & 40\,000 & 0.50 & SAX/LECS \\
  S2TL1.00  & 2500 & 1.00 & SAX/LECS \\
  S2TM1.00  & 10\,000 & 1.00 & SAX/LECS  \\
  S2TH1.00  & 40\,000 & 1.00 & SAX/LECS  \\[1pt]
\multicolumn{4}{c}{\em power-law EM source spectra} \\[1pt]
  APLM0.15 & 10\,000 & 0.15 & ASCA/SIS \\
  APLM0.50 & 10\,000 & 0.50 & ASCA/SIS \\
  APLM1.00 & 10\,000 & 1.00 & ASCA/SIS \\[1pt]
  SPLM0.15 & 10\,000 & 0.15 & SAX/LECS \\
  SPLM0.50 & 10\,000 & 0.50 & SAX/LECS \\
  SPLM1.00 & 10\,000 & 1.00 & SAX/LECS \\
\hline\\ [2pt]
\end{tabular}
\end{flushleft}
\end{table}

The XSPEC package has been used both to produce the simulated spectra
and to fit them. 300 realizations of the source spectrum with fixed
parameters and with a fixed total number of counts, including
statistical errors determined by Poisson statistics were generated.
The simulated spectra were then re-binned, so to have at least 20
counts in each resulting new (variable size) energy bin, to ensure
that the \cq\ statistics is applicable (\cite{edj+71}), and each
spectrum was fitted, always with a two-temperature {\sc mekal} model,
retrieving the best-fit spectral parameters.

The simulations assuming a two-temperature source spectrum were run
with 2500, 10\,000 and 40\,000 source counts, spanning the typical
source counts for actual SIS and LECS spectra. The simulations
assuming an intrinsic power-law spectrum were all performed with
10\,000 source counts. Simulations can supply at most a lower limit to
the uncertainty in the derived spectral parameters, because of at
least two optimistic assumptions which are only approximations to the
reality.  First, we are assuming that the {\sc mekal} plasma emission
code models describe perfectly the emission from a real plasma,
neglecting, among other things, all of the uncertainties in the atomic
physics.  Second, we are assuming that X-ray detectors are perfectly
calibrated, neglecting (possible) systematic uncertainties present in
the whole process.

Additionally, we are assuming that the two-temperature model or the
power-law emission measure distribution are adequate representations
of the source spectrum. On the basis of our knowledge of the solar
corona, as well as on the basis of analyses of EUV spectra of real
coronal sources, this is likely to be, at best, an approximate
parameterization of the coronae of real stars.

Abundance variations in the source spectrum have been modeled through
a single parameter, the global metallicity $Z$, thus assuming no
variations in the abundance ratios with respect to the solar values.
In fact, several of the analyses published on SIS spectra of coronal
sources result in best-fit spectra with significant variations in the
elemental abundance ratios. A simulation-based analysis for the case
of individually varying elemental abundances will be the subject of a
future paper.

Each numerical experiment is indicated in the following, with a code,
and the correspondence between the codes and the experiment parameters
is given in Table~\ref{tab:des}.

\section{Simulation results}

Each simulation yields a set of best-fit spectral parameters for all
the realizations of the sample spectrum. The distribution of the
best-fit values is indicative of the uncertainties that can be
expected, in the absence of systematic effects, when fitting real
spectra with similar statistics.

\begin{figure}[thbp]
  \begin{center}
    \leavevmode
%    \picplace{11.0 cm}
    \epsfig{file=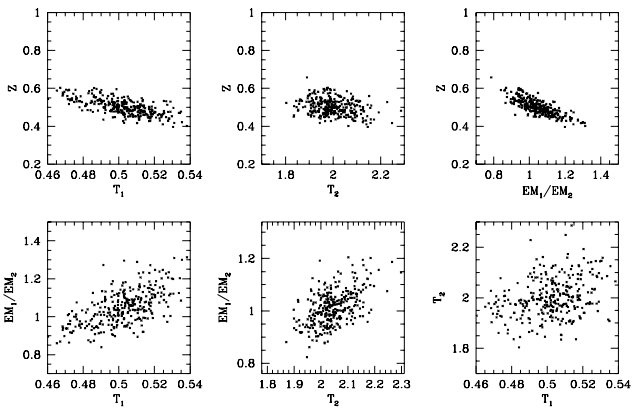, width=8.8cm, bbllx=20pt, bblly=175pt,
      bburx=580pt, bbury=530pt, clip=}
  \end{center}
  \caption{Two-temperature source spectra fit with two-temperature
    models: distribution of the best-fit parameters for a set of 300
    realizations, with coronal abundance 0.5 times solar as observed
    with the SAX/LECS detector, with 10\,000 counts accumulated per
    spectrum (experiment S2TM0.50).  }
  \label{fig:sax10k}
\end{figure}

\begin{figure}[htbp]
  \begin{center}
    \leavevmode
%    \picplace{11.0 cm}
    \epsfig{file=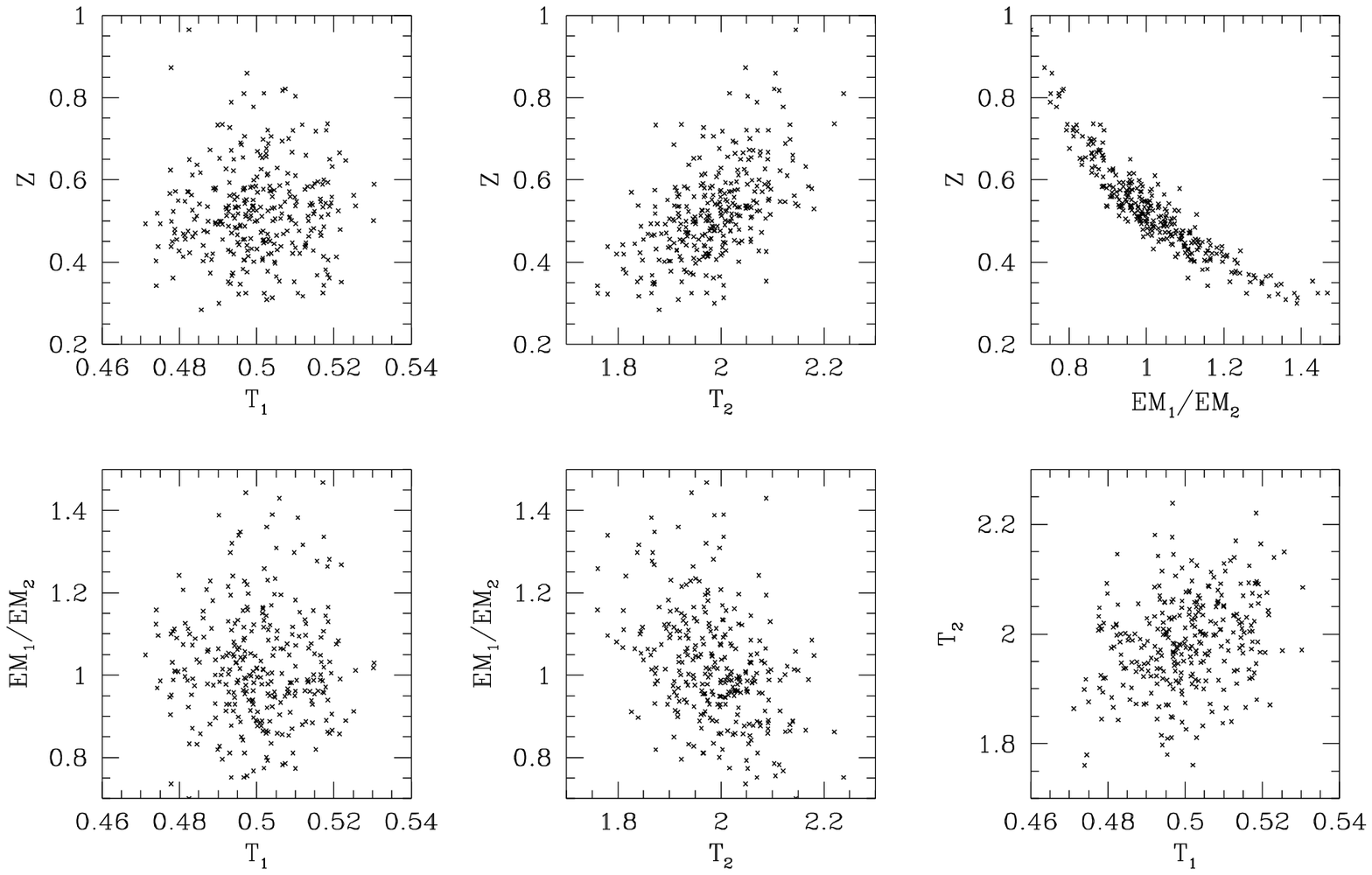, width=8.8cm, bbllx=20pt,
      bblly=175pt, bburx=580pt, bbury=530pt, clip=}
  \end{center}
  \caption{Two-temperature source spectra fit with two-temperature
    models: distribution of the best-fit parameters for a set of 300
    realizations, with coronal abundance 0.5 times solar as observed
    with the ASCA/SIS detector, with 10\,000 counts accumulated per
    spectrum (experiment A2TM0.50).  }
  \label{fig:asca10k}
\end{figure}

Some representative cases for the simulations assuming two-temperature
source spectra are shown in Figs.~\ref{fig:sax10k}
and~\ref{fig:asca10k}, in the form of scatter diagrams, plotting
best-fit spectral parameters against each other.  For each simulation
we have computed the median, as well as the upper and lower 68\% and
90\% quantiles, corresponding to ``1\,$\sigma$'' and ``2\,$\sigma$''
levels for each best-fit parameter. The median and quantiles are shown
in numerical form in Table~\ref{tab:quant}, and in graphical form in
Figs.~\ref{fig:efstat} and~\ref{fig:efz}.

\begin{table*}[htbp]
  \caption{Observed 1- and 2-$\sigma$ ranges for the best-fit
    parameters for all the numerical experiments discussed in the
    present paper. The entries in the table are in percent of the
    median value of the relevant best-fit parameter.}
\label{tab:quant}
\begin{flushleft}
%\scriptsize
  \footnotesize
\begin{tabular}{crrrrrrrrrr} \hline \\[-0.5pt]

Code & \multicolumn{2}{c}{$T_1$} & \multicolumn{2}{c}{$T_2$} &
 \multicolumn{2}{c}{$EM_1/EM_2$} &  \multicolumn{2}{c}{$Z$} \\
 & \multicolumn{1}{c}{1-$\sigma$} & \multicolumn{1}{c}{2-$\sigma$} &
 \multicolumn{1}{c}{1-$\sigma$} & 
 \multicolumn{1}{c}{2-$\sigma$} & \multicolumn{1}{c}{1-$\sigma$} &
 \multicolumn{1}{c}{2-$\sigma$} & \multicolumn{1}{c}{1-$\sigma$} &
 \multicolumn{1}{c}{2-$\sigma$} \\ 
\hline
  A2TL0.15  & $+7\hfill -7$ & $+13\hfill -15$ & $+13\hfill -6$ & $+29\hfill -12$ & $+74\hfill -29$ & $+175\hfill -40$ & $+71\hfill -44$ & $+114\hfill -65$ \\ 
  A2TM0.15  & $+3\hfill -3$ & $+5\hfill -5$ & $+2\hfill -4$ & $+5\hfill -7$ &      $+27\hfill -13$ & $+49\hfill -23$ & $+24\hfill -25$ & $+46\hfill -34$ \\   
  A2TH0.15  & $+1\hfill -1$ & $+3\hfill -3$ & $+0\hfill -3$ & $+1\hfill -4$ &      $+12\hfill -5$ & $+21\hfill -10$ & $+7\hfill -13$ & $+14\hfill -21$ \\     
  A2TL0.50  & $+4\hfill -4$ & $+7\hfill -8$ & $+11\hfill -7$ & $+18\hfill -11$ &   $+35\hfill -23$ & $+70\hfill -37$ & $+59\hfill -32$ & $+131\hfill -46$ \\  
  A2TM0.50  & $+2\hfill -2$ & $+3\hfill -4$ & $+3\hfill -4$ & $+6\hfill -7$ &      $+15\hfill -11$ & $+29\hfill -18$ & $+25\hfill -15$ & $+45\hfill -29$ \\   
  A2TH0.50  & $+1\hfill -1$ & $+1\hfill -2$ & $+0\hfill -3$ & $+1\hfill -4$ &      $+10\hfill -2$ & $+14\hfill -7$ & $+6\hfill -11$ & $+14\hfill -16$ \\      
  A2TL1.00  & $+2\hfill -2$ & $+4\hfill -4$ & $+17\hfill -7$ & $+26\hfill -14$ &   $+36\hfill -24$ & $+86\hfill -36$ & $+105\hfill -32$ & $+214\hfill -58$ \\ 
  A2TM1.00  & $+1\hfill -2$ & $+3\hfill -3$ & $+3\hfill -5$ & $+6\hfill -7$ &      $+17\hfill -9$ & $+27\hfill -16$ & $+28\hfill -21$ & $+52\hfill -31$ \\    
  A2TH1.00  & $+1\hfill -1$ & $+1\hfill -1$ & $+0\hfill -3$ & $+2\hfill -4$ &      $+9\hfill -3$ & $+14\hfill -8$ & $+10\hfill -11$ & $+21\hfill -18$ \\[5pt]      
  S2TL0.15   & $+11\hfill -11$ & $+16\hfill -20$ & $+13\hfill -5$ & $+20\hfill -11$ &$+37\hfill -15$ & $+58\hfill -28$ & $+21\hfill -19$ & $+58\hfill -28$ \\    
  S2TM0.15   & $+5\hfill -5$ & $+9\hfill -9$ & $+4\hfill -4$ & $+7\hfill -6$ &      $+18\hfill -6$ & $+25\hfill -13$ & $+10\hfill -9$ & $+19\hfill -15$ \\     
  S2TH0.15   & $+3\hfill -2$ & $+5\hfill -4$ & $+0\hfill -3$ & $+1\hfill -4$ &      $+6\hfill -4$ & $+12\hfill -7$ & $+5\hfill -5$ & $+9\hfill -8$ \\          
  S2TL0.50   & $+7\hfill -5$ & $+10\hfill -12$ & $+11\hfill -5$ & $+19\hfill -10$ & $+26\hfill -12$ & $+41\hfill -19$ & $+21\hfill -17$ & $+42\hfill -24$ \\   
  S2TM0.50   & $+3\hfill -2$ & $+5\hfill -5$ & $+4\hfill -3$ & $+6\hfill -5$ &      $+13\hfill -4$ & $+22\hfill -9$ & $+8\hfill -7$ & $+16\hfill -12$ \\       
  S2TH0.50   & $+1\hfill -1$ & $+2\hfill -3$ & $+0\hfill -2$ & $+2\hfill -3$ &      $+7\hfill -2$ & $+10\hfill -4$ & $+4\hfill -4$ & $+6\hfill -7$ \\          
  S2TL1.00   & $+5\hfill -5$ & $+8\hfill -9$ & $+10\hfill -6$ & $+18\hfill -10$ &   $+20\hfill -12$ & $+35\hfill -18$ & $+26\hfill -17$ & $+49\hfill -25$ \\   
  S2TM1.00   & $+2\hfill -2$ & $+4\hfill -4$ & $+3\hfill -3$ & $+6\hfill -5$ &      $+12\hfill -2$ & $+17\hfill -7$ & $+10\hfill -7$ & $+17\hfill -14$ \\      
  S2TH1.00   & $+1\hfill -1$ & $+9\hfill -7$ & $+3\hfill -3$ & $+5\hfill -5$ &      $+6\hfill -0$ & $+9\hfill -3$ & $+3\hfill -4$ & $+7\hfill -7$ \\[5pt]

  APLM0.15  & $+3\hfill -3$ & $+5\hfill -5$ & $+4\hfill -4$ & $+6\hfill -6$ &      $+9\hfill -8$ & $+15\hfill -13$ & $+16\hfill -14$ & $+29\hfill -21$ \\
  APLM0.50  & $+4\hfill -4$ & $+6\hfill -7$ & $+4\hfill -3$ & $+6\hfill -6$ &      $+15\hfill -10$ & $+23\hfill -14$ & $+14\hfill -15$ & $+29\hfill -24$ \\
  APLM1.00  & $+10\hfill -8$ & $+37\hfill -11$ & $+4\hfill -4$ & $+26\hfill -6$ &      $+36\hfill -22$ & $+403\hfill -30$ & $+22\hfill -24$ & $+33\hfill -60$ \\[5pt]

  SPLM0.15 & $+6\hfill -4$ & $+9\hfill -7$ & $+3\hfill -3$ & $+5\hfill -5$ &      $+18\hfill -13$ & $+33\hfill -21$ & $+8\hfill -9$ & $+14\hfill -15$ \\
  SPLM0.50 & $+6\hfill -7$ & $+13\hfill -11$ & $+4\hfill -3$ & $+9\hfill -5$ &      $+26\hfill -17$ & $+55\hfill -28$ & $+11\hfill -10$ & $+20\hfill -17$ \\
  SPLM1.00 & $+21\hfill -16$ & $+30\hfill -29$ & $+19\hfill -8$ & $+32\hfill -12$ &      $+130\hfill -43$ & $+234\hfill -59$ & $+33\hfill -19$ & $+57\hfill -28$ \\
\hline\\[2pt]
\end{tabular}
\end{flushleft}
\end{table*}

\subsection{Results for the SAX/LECS two-temperature simulations} 

Our simulations show that the combination of resolution and spectral
coverage offered by the LECS detector allows retrieval of the spectral
parameters through a \cq\ fit, for two-temperature spectra, with good
accuracy, already for spectra with $\simeq 10\,000$ counts The
2-$\sigma$ accuracy is $\simeq 5\%$ for the source temperatures, and
$\simeq 15\%$ for the coronal metallicity. LECS spectra with $\simeq
40\,000$ counts allow 2-$\sigma$ accuracies of $\simeq 2\%$ on the
source temperatures, and $\simeq 7\%$ on the coronal metallicity.

\begin{figure}[htbp]
  \begin{center}
    \leavevmode
%    \picplace{11.0 cm}
    \epsfig{file=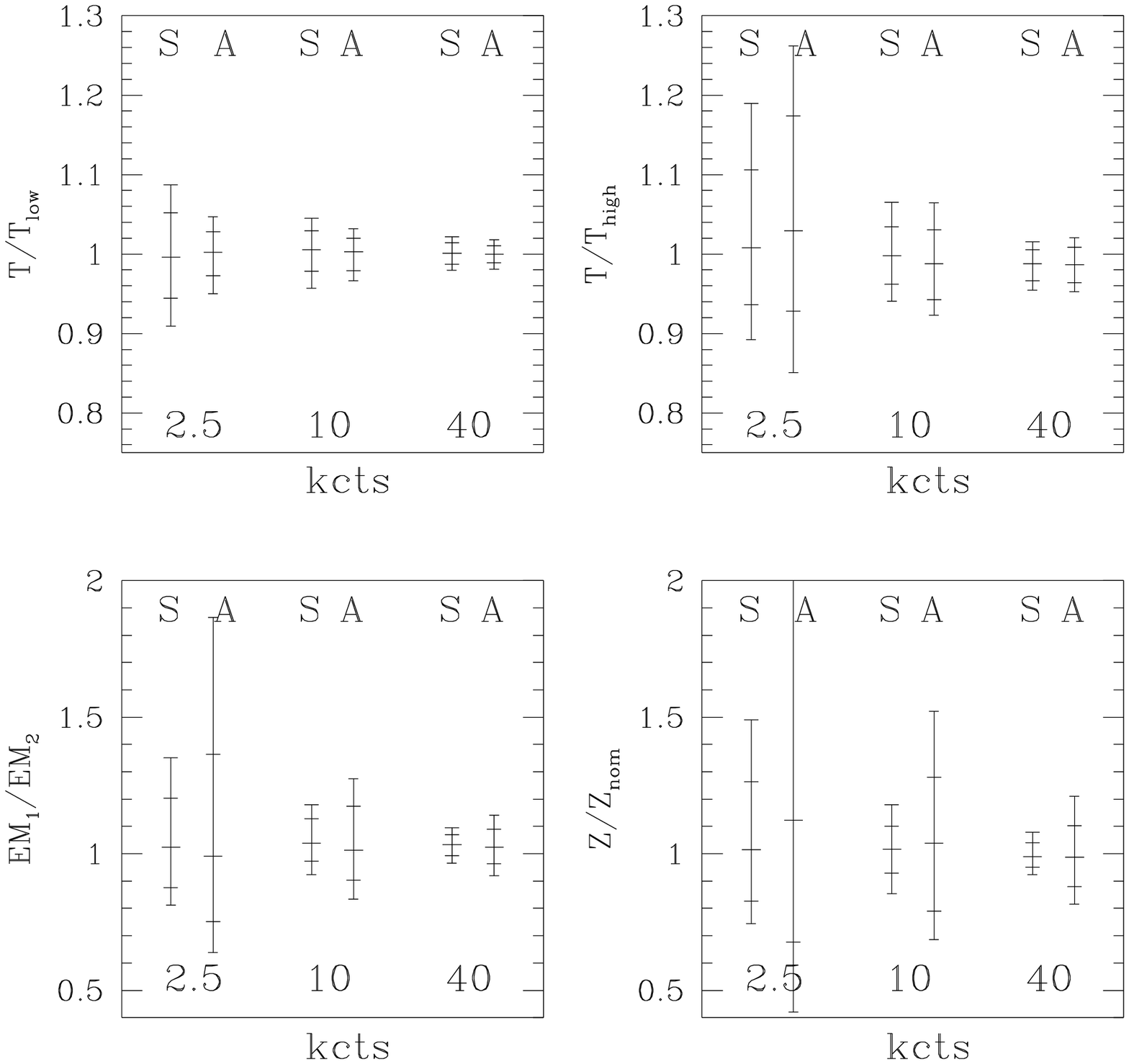, width=8.8cm, bbllx=0pt, bblly=175pt,
      bburx=580pt, bbury=700pt, clip=}
  \end{center}
  \caption{Two-temperature source spectra fit with two-temperature
    models: the effect of increasing photon statistics on the 68\% and
    90\% error ranges for the best-fit parameters. Each panel shows
    the error ranges for a given parameter, with the number of counts
    in the total spectrum indicated by the bottom label. The top label
    indicates the instrument (``S'' for SAX/LECS and ``A'' for
    ASCA/SIS). The two top panels show the error range for the cool
    (top left) and hot (top right) temperatures, while the bottom
    panels show the range for the ratio between the two emission
    measures (bottom left) and for the coronal metallicity (bottom
    right).}
  \label{fig:efstat}
\end{figure}

\begin{figure}[htbp]
  \begin{center}
    \leavevmode
%    \picplace{11.0 cm}
    \epsfig{file=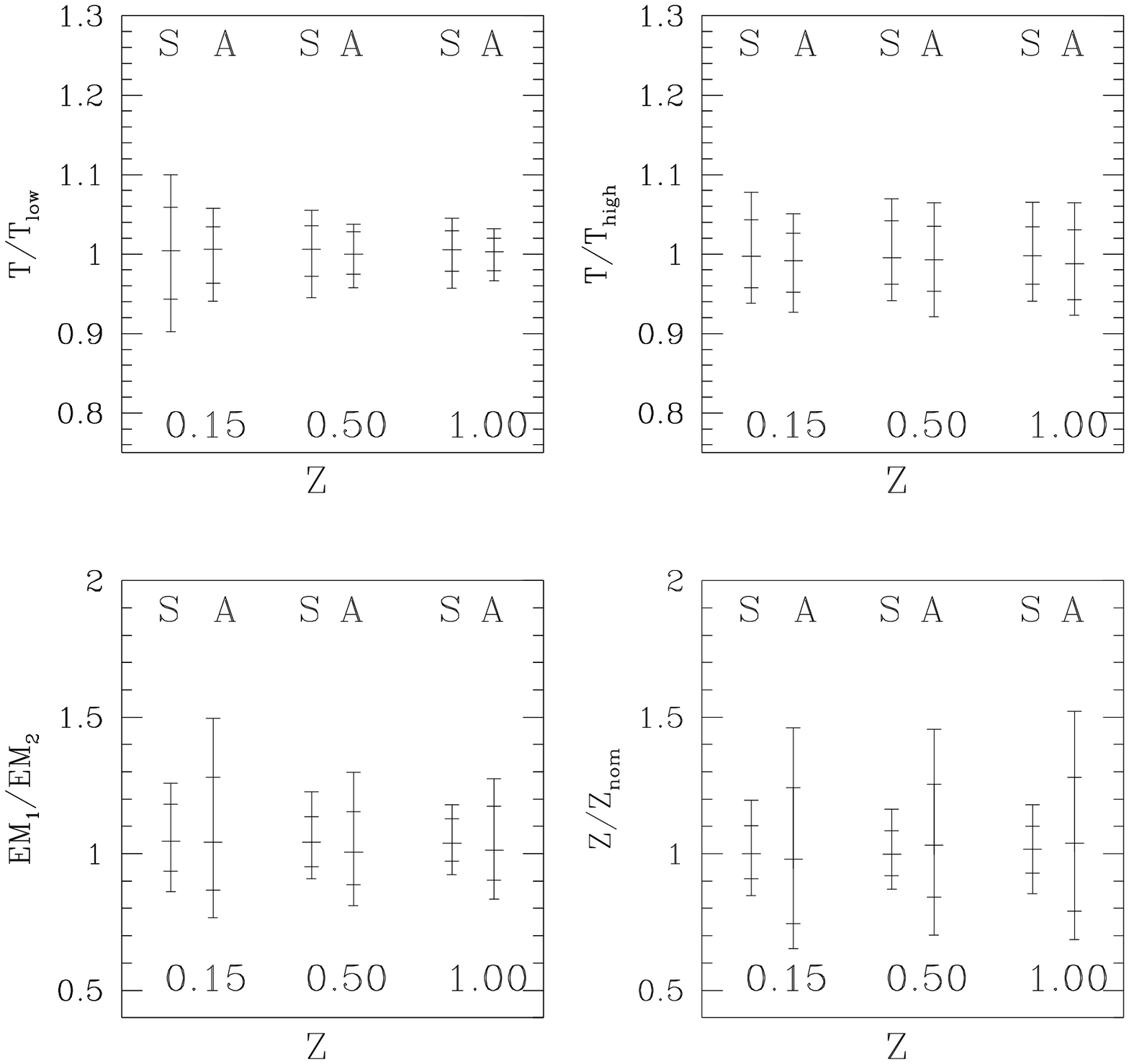, width=8.8cm, bbllx=0pt, bblly=175pt,
      bburx=580pt, bbury=700pt, clip=}
  \end{center}
  \caption{Two-temperature source spectra fit with two-temperature
    models: the effect of different source metallicity on the 68\% and
    90\% error ranges for the best-fit parameters, illustrated using
    the simulations with 10\,000 counts in the source spectrum. Each
    panel shows the error ranges for a given parameter, with the
    source metallicity indicated by the bottom label. The top label
    indicates the instrument (``S'' for SAX/LECS and ``A'' for
    ASCA/SIS). The panels have the same structure as in
    Fig.~\protect{\ref{fig:efstat}}.}
  \label{fig:efz}
\end{figure}

The best-fit parameters show some correlation (Fig.~\ref{fig:sax10k}),
i.e. they appear to be mildly dependent on each other in statistical
terms, but the range of variation of the metallicity vs. the emission
measure ratio or the temperatures is relatively small. The scatter
plots for the rest of the simulations are very similar to the ones in
Fig.~\ref{fig:sax10k}, with the same qualitative features, and a total
scatter decreasing as expected with increasing total number of counts,
as summarized in Fig.~\ref{fig:efstat}. At the same time, as shown in
Fig.~\ref{fig:efz}, the relative scatter is essentially independent on
the intrinsic source metallicity.

\subsection{Results for the ASCA/SIS two-temperature simulations}

The results for the simulations relative to the SIS detector differ
from the ones relative to the LECS, in the case of two-temperature
source spectra fit with two-temperature models, in one important way:
the scatter of the best-fit coronal metallicity value is (for the same
total number of counts in the spectrum) much larger, as it is the
scatter for the best-fit emission-measure ratio. The two are also
strongly correlated with each other.

The larger observed scatter implies that while the temperatures can be
measured in a $\simeq 10\,000$-counts SIS spectrum with a
2-$\sigma$ accuracy of $\simeq 5\%$, the coronal metallicity can
be measured, in the same spectrum, with a 2-$\sigma$ accuracy of only
$\simeq 40\%$, and an accuracy on the emission-measure ratio of
$\simeq 25\%$.  Even in the higher statistics spectra ($\simeq
40\,000$ counts), the 2-$\sigma$ accuracy for the source metallicity
is still only $\simeq 20\%$. This, again, in the optimistic
assumption of an intrinsic two-temperature source spectrum.

Again, the scatter plots not shown have the same qualitative features
of the ones shown, with the total scatter decreasing, as expected,
with increasing total number of counts.

\subsection{Results for the SAX/LECS power-law simulations}

\begin{figure}[thbp]
  \begin{center}
    \leavevmode
%    \picplace{11.0 cm}
    \epsfig{file=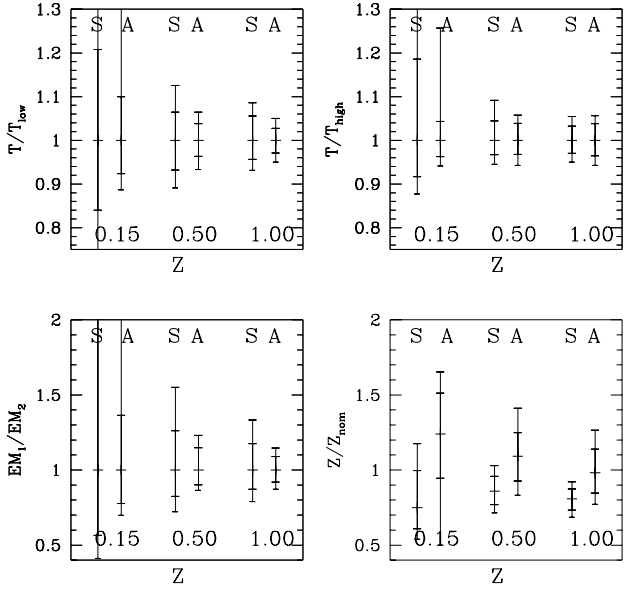, width=8.8cm, bbllx=0pt, bblly=175pt,
      bburx=580pt, bbury=700pt, clip=}
  \end{center}
  \caption{Power-law emission measure source spectra fit with two-temperature
    models: the effect of different source metallicity on the 68\% and
    90\% error ranges for the best-fit parameters.  Each panel shows
    the error ranges for a given parameter, with the source
    metallicity indicated by the bottom label. The top label indicates
    the instrument (``S'' for SAX/LECS and ``A'' for ASCA/SIS). The
    panels have the same structure as in
    Fig.~\protect{\ref{fig:efstat}}.}
  \label{fig:efplaw}
\end{figure}

The confidence ranges for the best-fit parameters obtained by fitting
a power-law source spectrum with a two-temperature model in the case
of the LECS detector are shown in Fig.~\ref{fig:efplaw}, together with
the results relative to the same type of simulations done for the SIS.
In this case the confidence ranges are not computed relative to the
``true'' source parameters (as the source model is different from the
model used to fit the data, and thus there are no ``true'' two
temperatures for the source) but rather with respect to the median
value found in the simulated set. The only parameter for which the
confidence regions are computed relative to the source value is the
metal abundance. Inspection of Fig.~\ref{fig:efplaw} and of
Table~\ref{tab:quant} shows that, if the intrinsic source metallicity
is really as low as 0.15 times solar, the confidence regions derived
by using a two-temperature model are exceedingly broad. We will
further discuss only the results relative to the $Z=1.0$ and $Z=0.5$
cases.

A comparison of Fig.~\ref{fig:efz} with Fig.~\ref{fig:efplaw} shows
that, for the LECS detector, the size of the confidence regions for
the best-fit temperatures has increased slightly in comparison with
the case of two-temperature source spectra, as it has for the best-fit
emission-measure ratio. The best-fit metallicity is also showing
slightly larger confidence regions, in addition to a significant
offset from its true value, with the two-temperature model retrieving
lower abundances than the true value.

\subsection{Results for the ASCA/SIS power-law simulations}

When the SIS is used to fit power-law source spectra with a
two-temperature model, the \cq\ on average does not converge to a
satisfactory value. For the cases with $Z=0.5$ and $Z=1.0$ the median
\cq\ values of the fit are 1.17 and 1.27, respectively, which on
account of the large number of \dof\ of the fits presented here
($\simeq 200$) corresponds to fits which are formally unacceptable
with a probability (for the $Z=1$ case) of $\simeq 0.5$\%. This has an
influence on the size of the confidence regions, and makes it appear
the fitting process to be ``more precise'' than it actually is. As a
result, the confidence regions for this case cannot be directly
compared with the previous cases, in all of which the best fit \cq\ 
distribution did not show any excess above the expected distribution.

The confidence regions for the SIS two-temperature fits of power-law
source spectra are still significantly larger than the LECS ones,
although they appear (because of the above-mentioned effect) to be
smaller than the regions for the two-temperature fitting of
two-temperature source spectra (again but for the case $Z=0.15$, which
shows an anomalous scatter). However, the median best-fit metallicity
appears to be close to the intrinsic metallicity of the original
power-law model.

\section{Comparison between the ASCA/SIS and SAX/LECS detectors}

The SIS and LECS detectors differ in their trade-offs of resolution
and spectral coverage, and thus behave differently in their capability
to retrieve source parameters, as well as in their sensitivity to
discrepancies between the intrinsic source spectral model and the
model being used to fit the data.

\subsection{Two-temperature intrinsic source spectra}
\label{sec:2tspec}

Given the same total number of counts in the spectrum, the SIS
detector does a slightly better job in determining the source
temperatures, but it performs worse than the LECS in terms of
retrieving the coronal metallicity of the source under investigation
(Figs.~\ref{fig:efstat} and~\ref{fig:efz}). A solar-metallicity,
$\simeq 10\,000$-counts LECS spectrum allows the coronal metallicity
to be determined with a 2-$\sigma$ accuracy of $\pm 15$\%, comparable
to the accuracy obtained with a $\simeq 40\,000$-counts SIS spectrum.

\begin{figure}[thbp]
  \begin{center}
    \leavevmode
%    \picplace{11.0 cm}
    \epsfig{file=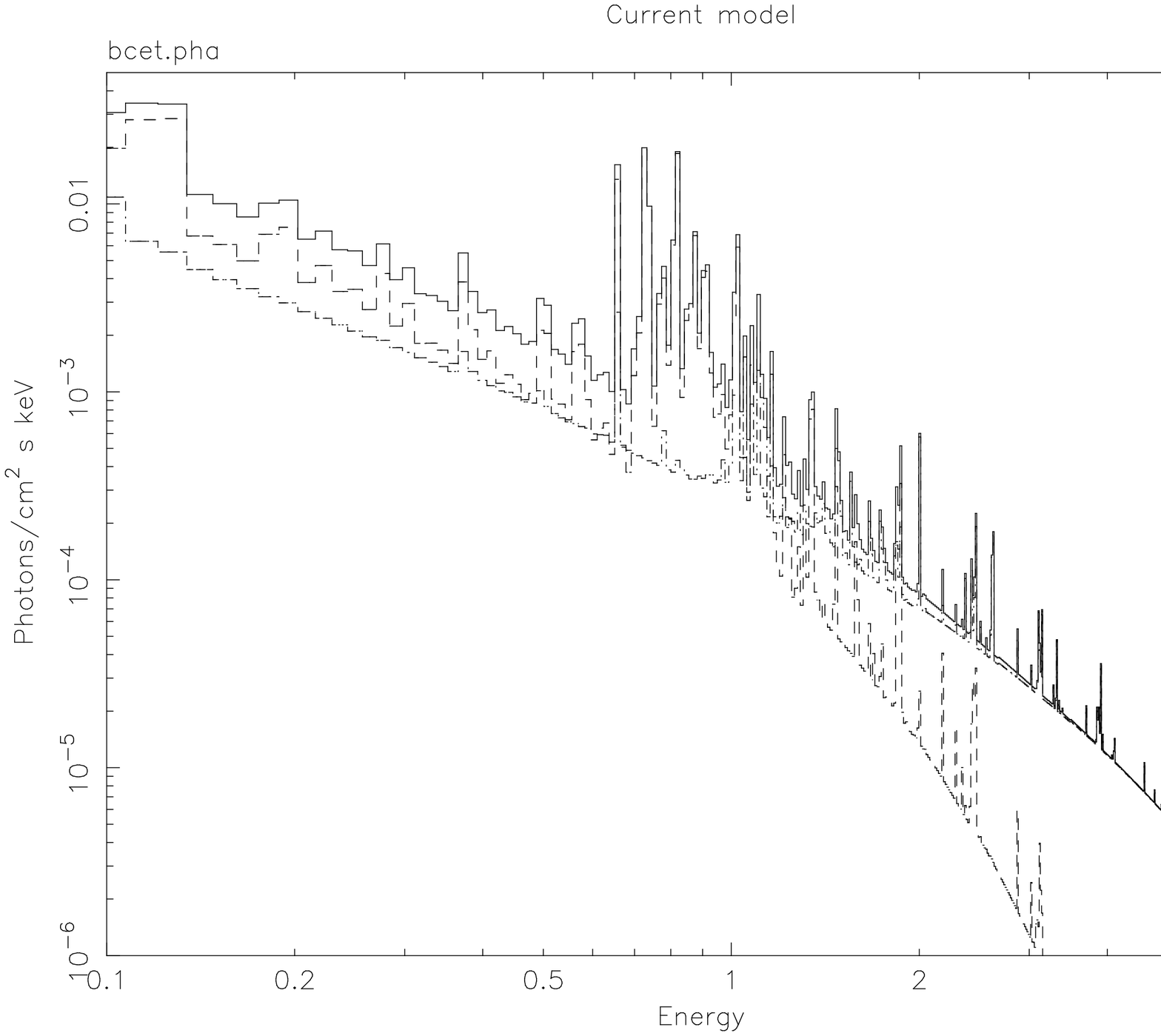, width=9.cm, bbllx=5pt, bblly=10pt,
      bburx=687pt, bbury=510pt, clip=}
  \end{center}
  \caption{The two-temperature model source spectrum, for solar
    abundance, used in the present work. Notice the almost line-free
    continuum below 0.5\,keV, which can be exploited by the SAX/LECS
    for metallicity determinations and which is not available to the
    ASCA/SIS. The two dashed lines are the two individual temperature
    components, while the continuos line is the total spectrum.}
  \label{fig:model}
\end{figure}

The reason for this large difference lies in the diagnostic which is
being exploited by the fitting process to derive the abundance.  The
metallicity is essentially determined by balancing the spectrum in the
region between 0.7 and 1.2\,keV, where line emission (mostly by Fe\,L
lines) dominates and where the statistics are best, both in the SIS
and in the LECS, with the available continuum emission. The region
around 1\,keV is thus acting as a sort of ``pivot'' for the fitting
process.  In the case of the SIS, little continuum is usually
available, mostly in the (limited) region between 0.5 and 0.7\,keV, in
which however the detector's resolution is lowest, and thus the
continuum contribution may not be easily disentangled from the nearby
line emission. The continuum in the hard tail of the spectrum,
typically the small regions between the major K-complexes of heavier
elements has, for most coronal sources, low intrinsic source flux, and
thus low signal to noise ratio.

\begin{figure}[htbp]
  \begin{center}
    \leavevmode
%    \picplace{11.0 cm}
    \epsfig{file=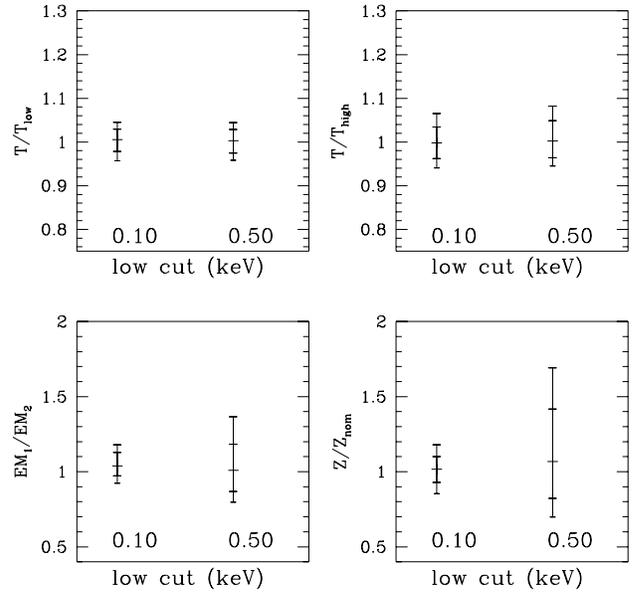, width=8.8cm, bbllx=0pt, bblly=175pt,
      bburx=580pt, bbury=700pt, clip=}
  \end{center}
  \caption{Two-temperature source spectra fit with two-temperature
    models: the effect of moving the low-energy cut for SAX/LECS
    spectra from 0.1 to 0.5\,keV is illustrated for the sample case of
    solar-metallicity spectra with 10\,000 counts. The panels have the
    same structure as in Fig.~\protect{\ref{fig:efstat}}.}
  \label{fig:efcut}
\end{figure}

For LECS spectra the region below 0.5\,keV supplies a very well
constrained determination of the continuum, and makes the fitting
process much more robust. To confirm the importance of the region
below 0.5\,keV we have performed an additional set of simulations for
the LECS, fitting only the spectrum above 0.5\,keV, i.e.\ using only
the information available to the SIS detector. As expected, the
performance for the LECS becomes comparable to the performance of the
SIS (actually slightly worse, not surprisingly, given its lower
spectral resolution). Fig.~\ref{fig:efcut} shows the effect of moving
the low-energy cut from 0.1 to 0.5\,keV for LECS spectra on the
confidence regions, for the sample case of solar-metallicity spectra
with 10\,000 counts.

\subsection{Power-law intrinsic source spectra}

When power-law source spectra are fit with two-temperature models, the
two detectors discussed here behave differently. While the LECS
produces formally acceptable fits, but with lower best-fit
metallicities than the intrinsic source value, the SIS detector
produce formally unacceptable fits, while still converging on the
right source metallicity.

This behavior can be explained as follows: when performing simulations
with the same total number of counts for the SIS and the LECS, the
smaller energy coverage of the SIS results in a concentration of
counts in the $\simeq 1$\,keV region, yielding spectra with higher
statistics than in the case of the LECS, in which the counts are
spread in a larger spectral region with two distinct peaks (at $\simeq
0.2$ and $\simeq 1$\,keV), so that the resulting statistics
(particularly in the peak channels) are correspondingly lower. This
explains why the SIS spectra in this case result, with the same number
of total counts, in worse \cq\ values than the LECS ones.

The convergence of LECS spectra on a lower than intrinsic metallicity
value is due to a different effect: LECS fits are driven mainly by the
balance between the two high-statistics regions of the spectrum, i.e.
the continuum below $\simeq 0.4$\,keV and the line-rich region around
$\simeq 1$\,keV. When attempting to fit power-law spectra with
two-temperature models the discrepancy between the source spectrum and
the fitting model is such that the fit is lead to a lower abundance
model by the need to balance the lines and the continuum. This does
not happen in the case of the SIS, again because of the lack of
high-statistics continuum regions.

\begin{figure}[htbp]
  \begin{center}
    \leavevmode
%    \picplace{11.0 cm}
    \epsfig{file=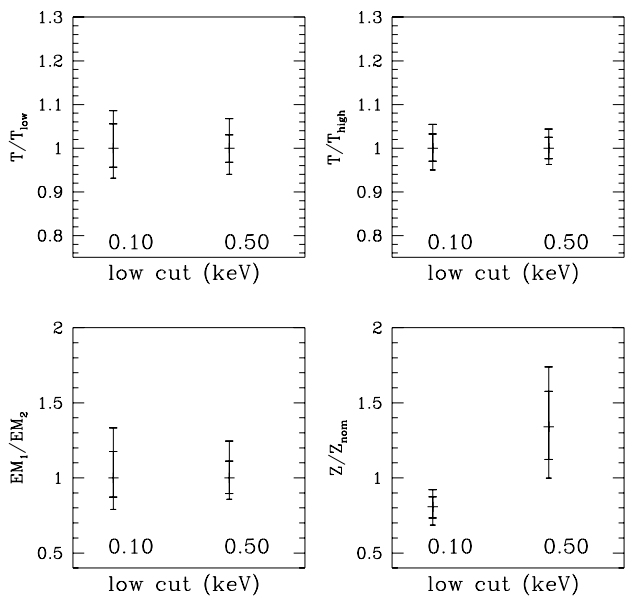, width=8.8cm, bbllx=0pt, bblly=175pt,
      bburx=580pt, bbury=700pt, clip=}
  \end{center}
  \caption{Power-law emission measure source spectra fit with two-temperature
    models: the effect of moving the low-energy cut for SAX/LECS
    spectra from 0.1 to 0.5\,keV is illustrated for the sample case of
    solar-metallicity spectra with 10\,000 counts. }
  \label{fig:efcutplaw}
\end{figure}

The reality of the above effect is well shown by a further set of
simulations, in which the same LECS power-law source spectra are fit
only using the spectral region above 0.5\,keV. In this case the fit is
``pivoting'' around the high-statistics 1\,keV region, and is being
driven to higher metallicities (this time higher than the intrinsic
source metallicity) by the residuals in the high-energy tail.
Fig.~\ref{fig:efcutplaw} shows how the confidence regions change when
the soft cut in the LECS spectra is moved from 0.1 to 0.5\,keV,
indicating that when the fitting model is not strictly adhering to the
true source thermal distribution the best-fit metallicity can be
strongly influenced by the spectral region being used.  Conversely, a
best-fit metallicity which changes with the spectral region being fit
is simptomatic of too simplistic a model. The direction and the size
of the effect in real cases will depend on the (a priori unknown)
intrinsic source temperature distribution, so that the present case
only has an illustrative value.

\section{Two test cases: the observed ASCA/SIS spectra of $\beta$~Ceti 
and of Capella}
\label{sec:asca}

We have analyzed the SIS spectra of the active giant \bcet\ and of the
active binary Capella obtained during the Performance Verification
(PV) phase. The data were retrieved from the public archive, and
analyzed according to the procedures suggested for ASCA data.

The data for \bcet\ consist of two segments, of $\simeq 8$ and $\simeq
3.4$\,ks each, taken in two different instrumental modes ({\sc faint}
and {\sc bright} mode).  They have thus been processed separately, but
have been fit simultaneously, as discussed in \cite*{mfp+97}. Although
the \bcet\ SIS spectrum is better fit with a model in which the
individual abundances are left free to vary (as is the Capella
spectrum), for coherence with the simulations, they have been fit here
with global varying abundances.  Again, we have used the {\sc mekal}
emission code as implemented in XSPEC V.9.0; the results therefore are
not directly comparable with the ones of \cite*{dsw+94}, obtained with
the {\sc meka} emission code (an earlier version of the {\sc mekal}
code, see \cite{mkl95} for a description of the changes).  Eventual
differences in the version of the ASCA pipeline processing may also
influence the results. The two resulting spectra have been rebinned at
a minimum of 20\,cts per bin.  The summed number of counts from both
spectra was $\simeq 10\,000$.

The \bcet\ SIS spectrum was first fitted across a broad band, from
0.5\,keV up to 4.0\,keV. In this energy range the best-fit metallicity
is $\simeq 0.9$ times solar. However, the exclusion of channels at the
soft end of the spectrum causes the best-fit metallicity to change,
making it decrease to 0.33 solar if channels with energies from 0.5 to
0.6\,keV are excluded, and as low as 0.22 if channels up to 0.7\,keV
are excluded.  At the same time, the ratio between the emission
measures of the cool and of the hot component increases, showing the
same type of behavior as the simulations for the case of intrinsic
two-temperature source spectra, where lower best-fit metallicities are
correlated with higher emission-measure ratios.

The same type of analysis was done for the Capella spectrum, finding a
best-fit abundance for the broad-band (0.5--5.0\,keV) SIS Capella PV
phase spectrum of 0.54 times solar. As for \bcet, the exclusion of
spectral channels below 0.6\,keV lowers the best-fit abundance to 0.24
times solar, and exclusion of the channels below 0.7\,keV lowers it
even more, down to 0.18 times solar. As for \bcet\, these results are
not directly comparable to the ones of \cite*{dra96}.

The \cq\ does not converge, neither for Capella, nor for \bcet\, to
formally acceptable values. The physical meaning of best-fit
parameters is thus clearly questionable, and they should not therefore
be taken at face value. The strong and systematic dependence of the
best-fit metallicity on the spectral region in the Capella and \bcet\ 
spectra indicates the presence of some systematic error either in the
spectral model used (i.e. the emission code itself, the source
temperature structure, the assumption of solar metallicity ratio or a
combination of these) or in the detector's calibration (or both).

The continuum region used by the SIS fits (the region between $\simeq
0.5$ and $\simeq 0.75$\,keV) is almost (but not entirely) line-free,
but, at the resolution of the SIS, and given the calibration
uncertainty in the region (\cite{dmy+96}) it contains a significant
number of source counts ``spilling over'' from the Fe-L complex. The
statistics in this region are, for coronal sources, invariably higher
than for the hard tail where line complex from the K-transitions of
the heavier elements are visible, and will accordingly influence the
metallicity determination.

\begin{figure}[htbp]
  \begin{center}
    \leavevmode
%    \picplace{11.0 cm}
    \epsfig{file=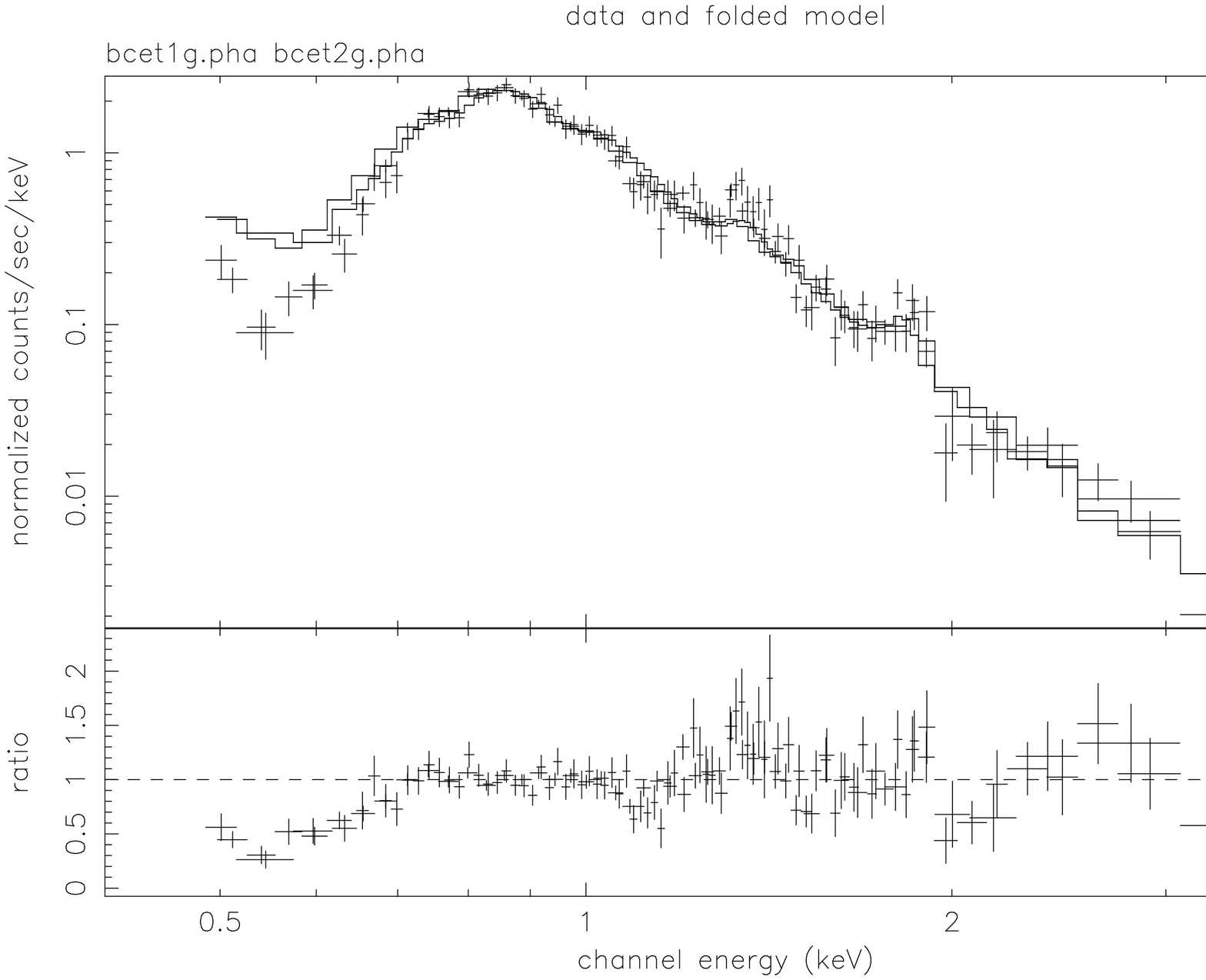, width=9.cm, bbllx=5pt, bblly=30pt,
      bburx=687pt, bbury=475pt, clip=}
  \end{center}
  \caption{The ASCA/SIS \bcet\ spectra, together with the best-fit
    model determined only considering the spectral range above
    0.7\,keV. Note that this model significantly over-shoots, as
    described in the text, the data in the softer region
    0.5--0.7\,keV.}
  \label{fig:bcet}
\end{figure}

The increase of the best-fit metallicity with inclusion of a softer
spectral region can be explained by assuming that the models being
used for the fit (either because of problems in the emissivity models
or because of calibration problems, or a mixture of both) have a lack
of flux in the region around 1\,keV (which is dominated by the cooler
temperature component). This could happen, for example, if a number of
lines which exist in real spectra were missing from the models. In
this case the fit process would converge by balancing the apparent
flux deficit around the line-rich 1\,keV region. The strategy
``chosen'' by the fit process will be different depending on the
spectral region being fit. If the softer channels are excluded, the
fit process can freely increase the cool emission measure to drive up
the 1\,keV region, while at the same time decreasing the metallicity
to avoid discrepancy in the hard tail, dominated by the hot component.
If the softer channels are included, however, the high emission
measure of the cool component drives up the pseudo-continuum around
0.5\,keV, and thus the fit process has to increase the metallicity (to
add flux around 1\,keV without adding more continuum at 0.5\,keV) and
correspondingly decrease the emission measure of the soft component.
The effect is well visible in the SIS spectrum of \bcet\ shown in
Fig.~\ref{fig:bcet}.

We have also fit the same spectra ignoring the region between 0.75 and
1.1\,keV. The result reinforces the above conclusions: the best-fit
metallicity for the \bcet\ spectrum rises to 1.1 times solar, the same
value which is found for the Capella spectrum applying the same
procedure. The lower metallicity obtained by fitting the whole
spectrum could be explained with a lack of flux (again due to
calibration effects and/or to lack of lines in the emission models)
around the 1\,keV region. Another possibility (which is not mutually
exclusive with the first one) is that the two-temperature model being
used to fit the data is not a good representation of the real
temperature distribution of the emitting plasma.

The possible lack of lines in current plasma emission codes in the
region around 1\,keV has for example been discussed by \cite*{jor96},
who makes explicit reference to the {\sc meka} model (from which the
{\sc mekal} model used here has been derived by improving the Fe
ionization balance and by adding a number of lines), concluding that
{\em ``in view of the large number of weak transitions not yet
  included in the emissivity codes it seems premature to attribute
  discrepancies between the observations and the predictions of the
  codes to non-photospheric abundances''}.

The dependence of derived abundances on the precise spectral interval
being fit does {\em not} disappear if a model with all the abundances
free to vary is used. Rather, it becomes more complicated because of
the interplay between the various abundances, and because introducing
a cut-off in the soft spectral region will make some specific
abundances impossible to determine (notably nitrogen and oxygen).
However, the resulting metallicities still strongly depend on the
spectral interval being considered.

\section{Discussion}

The set of simulations presented here allow us to get some insight in
the complex process of determining coronal abundances by global
fitting of X-ray spectra, as well as to determine the expected range
of uncertainty for the best-fit parameters as a function of spectrum
statistics and of the intrinsic source metallicity. In addition, by
giving an estimate of the range of variation in best-fit parameters
that can be induced by a certain level of {\em statistical} error, it
allows to estimate what can be the (approximate) effect of a certain
level of {\em systematic} error present, for example, in the detector
calibration or in the emissivity model.

The LECS performs better in constraning the intrinsic source
metallicity than the SIS. This is due to the larger spectral range
accessible, and in particular to the availability of a high
signal-to-noise continuum region below 0.5\,keV, with few and
relatively weak lines, which appears to be more important, for
spectroscopy of coronal sources, than the enhanced resolution of the
SIS.  In fact, as reported in Sect. \ref{sec:2tspec}, when we remove
the spectral region below 0.5 keV from the LECS spectra, thus making
its passband equal to the one of the SIS, the indetermination on the
metallicity derived from spectra of the two instruments is similar, as
it is the strong correlation between metallicity and emission measure
ratio.

Some results from the simulations are remarkably different from what
could have been expected. For example the cool component of the
two-temperature source models is better constrained by the SIS than by
the LECS (see Fig.~\ref{fig:efstat}). Naively, one would have expected
that the softer response of the LECS should have led to smaller
uncertainty on the cooler temperature. This example shows how the
complex interactions of parameters in the fit process can lead to
non-intuitive behavior and points, a posteriori, to the importance of
basing the fit process on detailed simulation work.

Simulations can only provide a limited insight as they are done on the
assumption that both the detector's calibration and the emissivity
model are not affected by significant errors. However, they also show
that even relatively small amounts of systematic errors can have a
strong influence in the derived best-fit parameters. The statistical
error of a 10\,000 counts coronal spectrum in the peak (i.e. around
1\,keV) is $\simeq 2.5$\% per effective resolution element (obtained
adding up the number of counts in the energy range corresponding to
the FWHM resolution at 1\,keV), for both the LECS and the SIS.  This
level of statistical uncertainty is sufficient to induce a 50\% spread
in the best-fit metallicity for SIS data at the 90\% confidence level.
Thus, similar levels of systematic errors, if present in the regions
in which the spectrum has a high signal-to-noise, could induce
comparable {\em systematic}\/ shifts in the best-fit metallicity.

The analysis of real SIS spectra done in Sect.~\ref{sec:asca} points
toward the likely presence of systematic effects due either to
calibration uncertainties or to problems in the emissivity models or
to the simplistic model used to fit the data (or to a combination of
all the above effects), as shown by the strong dependence of the
best-fit metallicity on the energy range used in the fit. In
particular, our results on the soft sources \bcet\ and Capella could
indicate a lack of predicted model flux in the region around 1\,keV.

The present results point toward several items which should be
considered critically when discussing coronal metallicities derived
from low-resolution X-ray spectra. The most important are:

\begin{itemize}
  
\item Calibration uncertainties are likely to induce systematic
  effects in the determination of coronal abundances. As shown above,
  calibration uncertainties of a few percent could induce systematic
  shifts in the best-fit abundances of several tens of percent.
  Calibration uncertainties in the spectral region in which the
  signal-to-noise of stellar spectra is higher (around 1\,keV) are
  potentially the most detrimental. Future instruments, such as XMM
  and AXAF, should thus pay particular attention to calibration in
  this spectral region (as well as in the region below 0.5\,keV).

\item Uncertainties in the plasma emissivity codes are as critical as
  detector calibration issues. Again, the region around 1\,keV is the
  one which should be the made as reliable as possible. Flux deficits
  in this region will have a very strong influence in the derived
  source parameters, not in terms of larger confidence regions, but
  rather in terms of systematic shifts of best-fit values, due to the
  pivoting role of the region of the spectrum with the highest
  signal-to-noise ratio.
  
  This becomes even more critical if individual abundances are left
  free to vary, as elements like Ca and Ar, which have their
  K-complexes in the hard region of the spectrum, have their L-complex
  lines also in the $\simeq 1$\,keV region. Again, the fit process
  will be more influenced by the higher signal-to-noise L lines than
  by the better resolved but noisier K lines. Given the strong
  blending in the 1\,keV region, uncertainties in few of the lines
  have a ripple-down effect on all abundances determined in the
  spectrum.

\item Assuming too simplistic a model for the fit (as illustrated here
  by fitting power-law temperature distributions with two-temperature
  models) can also induce systematic shifts in the retrieved
  abundance. As the evidence from EUVE spectral analyses of coronal
  sources points toward complex plasma temperature distributions, it
  is not unlikely that two-temperature analyses of real coronal
  spectra are affected by similar effects. In this regard the LECS
  wide energy range provides improved diagnostic capabilities, thanks
  to the presence of two pivoting regions, one which is line-rich and
  one which is almost line-free.

  A hint of the possible inadequacy of the spectral model being used
  is, in the case of the LECS, the derivation of different
  best-fit abundances as a function of the adopted low-energy
  cut-off. As shown by the fit of the SIS data of Capella and
  \bcet\ discussed here, this may also be true with ASCA, although
  calibration uncertainties in the softer detector channels may also
  be playing a role there.

\end{itemize}

\acknowledgements{We would like to thank A.\,N. Parmar and N.\,S.
  Brickhouse for the useful discussions related to the subject of the
  present work. A.\,M., G.\,P. and S.\,S. acknowledge financial
  support from ASI (Agenzia Spaziale Italiana), MURST (Ministero della
  Universit\`a e della Ricerca Scientifica e Tecnologica), and GNA-CNR
  (Gruppo Nazionale Astronomia del Consiglio Nazionale delle
  Ricerche).}

%\bibliographystyle{aabib}
%\bibliography{references}

\end{document}